\documentclass[twocolumn]{aastex63}
\usepackage{amsmath}
\usepackage{comment}

\newcommand{\vdiff}{v_{\rm diff}}

\newcommand{\lcool}{\lambda_{\rm F,turb}}
\newcommand{\tcool}{t_{\rm cool}}
\newcommand{\ttcool}{\tilde{t}_{\rm cool}}

\newcommand{\tcoolmin}{t_{\rm cool, min}}
\newcommand{\tmix}{t_{\rm cool, mix}}
\newcommand{\edot}{\dot{E}_{\rm cool}}
\newcommand{\edotc}{\dot{\varepsilon}_{\rm cool}}

\newcommand{\edoth}{\dot{\varepsilon}_{\rm heat}}
\newcommand{\tsh}{t_{\rm sh}}
\newcommand{\Lint}{L_{\rm int}}
\newcommand{\Lf}{L_{\rm frac}}
\newcommand{\vtL}{v_t(\Lint)}

\newcommand{\vtl}{v_t(\ell)}
\newcommand{\teddy}{t_{\rm eddy}}
\newcommand{\Lbox}{L_{\rm box}}

\newcommand{\Da}{{\rm Da}}
\newcommand{\Dasim}{{\rm Da}_{\rm sim}}

\newcommand{\treact}{t_{\rm react}}

\newcommand{\lT}{\lambda_T}
\newcommand{\lK}{\lambda_K}

\newcommand{\lFt}{\lambda_{\rm cool}}

\newcommand{\rhobar}{\rho_0}
\newcommand{\vrel}{v_{\rm rel}}
\newcommand{\mach}{\mathcal{M}}
\newcommand{\Pbar}{P_0}

\newcommand{\cshot}{c_{s,{\rm hot}}}

\newcommand{\Tcold}{T_{\rm cold}}
\newcommand{\Tpk}{T_{\rm pk}}
\newcommand{\Thot}{T_{\rm hot}}
\newcommand{\Tmix}{T_{\rm mix}}

\newcommand{\bhi}{\beta_{\rm hi}}
\newcommand{\blo}{\beta_{\rm lo}}

\newcommand{\Aint}{A_{\rm int}}

\newcommand{\vvec}{\mathbf{v}}
\newcommand{\vinloc}{v_{\rm in}}

\newcommand{\vbulk}{v_{\rm bulk}}

\newcommand{\dx}{\Delta x}
\newcommand{\nres}{N_{\rm res}}

\newcommand{\ak}{\texttt{AthenaK}\ }

\newcommand{\paperi}{Paper 1}

\received{XXX}
\revised{XXX}
\accepted{XXX}

\submitjournal{ApJ}

\shorttitle{The Origin of Da Scaling}
\shortauthors{Lancaster et al.}

\begin{document}

\title{The Origin of Da Scaling: Suppressed Cooling in Fast-Cooling Mixing Layers}

\correspondingauthor{Lachlan Lancaster}
\email{llancaster@flatironinstitute.org}

\author[0000-0002-0041-4356]{Lachlan Lancaster}
\thanks{Simons Fellow}
\affiliation{Center for Computational Astrophysics, Flatiron Institute, 162 5th Avenue, New York, NY 10010, USA}
\affiliation{Department of Astronomy, Columbia University,  550 W 120th St, New York, NY 10025, USA}

\author[0000-0003-3806-8548]{Drummond Fielding}
\affiliation{Department of Physics, New York University, 726 Broadway, New York, NY, 10003, USA}

\author[0000-0002-1600-7552]{Rajsekhar Mohapatra}
\affiliation{Department of Astrophysical Sciences, Princeton University, Princeton, NJ 08544, USA}

\author[0000-0003-2630-9228]{Greg L. Bryan}
\affiliation{Department of Astronomy, Columbia University,  550 W 120th St, New York, NY 10025, USA}
\affiliation{Center for Computational Astrophysics, Flatiron Institute, 162 5th Avenue, New York, NY 10010, USA}

\begin{abstract}
In numerical experiments simulating Turbulent Radiative Mixing Layers (TRMLs) it is observed that as the cooling time in the mixed gas, $\tcool$, becomes very short compared to the dynamical time of the turbulence, $\teddy/\tcool \gg 1$, there is a change in the scaling behavior of the total energy radiated in the TRML as a function of this ratio, also known as the Damk\"{o}hler number, $\Da \equiv \teddy/\tcool$, from $\edot \propto \Da^{1/2}$ to $\edot \propto \Da^{1/4}$. The latter, so-called ``fast-cooling,'' regime is of particular interest as many astrophysical mixing layers lie in this regime. We demonstrate that the origin of this change is the suppression of turbulent folding of the surface by the ram-pressure of the inflowing gas, which becomes much greater than the turbulent pressure in this regime. We present an argument that reproduces the $\edot \propto \Da^{1/4}$ behavior by appealing to the suppression of the fractal structure of the interface by the ram-pressure of the inflowing gas.
\end{abstract}


\section{Introduction}
\label{sec:intro}

Astrophysical fluids are ubiquitously turbulent and multiphase \citep{Field65,CowieMcKee77,BegelmanFabian90,Roepke07,KimOstrikerRaileanu17,Mandelker20,TonnesenBryan21,Lancaster21b,FB22,FGO23,Mohapatra25}. The ubiquity of turbulence means mixing is usually the dominant mechanism by which energy and momentum are communicated between these phases. The most abundant phases in a mass-, volume-, or energy-weighted sense are thermally stable\footnote{At least informally, as in the case of the hot phase there is no well-understood volumetric heating term that makes the phase stable but energy can be replenished hydro-dynamically (e.g. through feedback).} on timescales comparable to the dynamical time of the system, but the intermediate phase gas obtained by mixing at the interface between these phases often is not \citep{KK13,JenningsLi21,GronkeOh18,Abbott82}. In this work we will only consider the case where energy is lost radiatively (through the emission of light) though there are circumstances under which energy can be gained in the layer \citep{ZDPN69}. Focusing our attention on a section of the interface between two phases of astrophysical gas, we find a region whose dynamics are dominated by turbulent mixing and the radiative cooling process that makes the intermediate phase thermally unstable: a Turbulent Radiative Mixing Layer (TRML).

The TRML system is therefore a fundamental building block for understanding the transport of energy in astrophysical fluids. These systems are characterized by (i) the turbulence in the layer, (ii) the hydrodynamic properties of the two phases, such as their temperature and density, and (iii) the details of the radiative processes that take place in the mixed material. We will generally describe the turbulence in the layer by the (square-root of the) second-order structure function of the velocity field on scale, $\ell$
\begin{equation}
    \label{eq:vsf2_def}
    \vtl = \left[\frac{1}{V}\int_V \left\langle|\vvec(\mathbf{r}) - \vvec(\mathbf{r} + \mathbf{\delta})|^2 \right\rangle_{|\mathbf{\delta}| = \ell} d\mathbf{r} \right]^{1/2} \, .
\end{equation}
Where the calculation is carried out in some volume $V$ limited to the mixing layer. This is the same as the root-mean-square (r.m.s) velocity fluctuation on length scale $\ell$. This function will generally peak at a given scale, $\Lint$, called the integral scale, and $\vtL$ is on the order of the r.m.s. velocity in the whole volume $V$. The two phases are generally assumed to be in pressure equilibrium (as dynamical times are generally short compared to sound crossing times, at least in the hot phase), so that their difference in density and pressure are both determined by a single parameter
\begin{equation}
    \label{eq:chi_def}
    \chi \equiv \frac{\rho_{\rm cold}}{\rho_{\rm hot}}
\end{equation}
where $\rho_{\rm cold}$ is the density of the higher density phase (and therefore colder by pressure balance) and $\rho_{\rm hot}$ the smaller density, so that $\chi > 1$. Cooling in the mixed gas is assumed to occur on some characteristic timescale, $\tcool$ (a discussion of the correct time-scale to choose in terms of the full cooling function is given in \autoref{app:Da_def} and \autoref{app:tcool}).

In a steady-state, cooling in the layer is balanced by a flux of high specific-entropy gas from the hot phase into the mixing layer. Considering this flux through some interface associated with the mixing layer (we will specify below) with area $\Aint$ we can write the total cooling as
\begin{equation}
    \label{eq:edot_generic}
    \edot = \frac{\gamma}{\gamma -1} P \vinloc \Aint
\end{equation}
where $\gamma$ is the adiabatic index, $P$ is the pressure of the gas, and $\vinloc$ is the surface-averaged velocity at which gas is moved into the layer across the surface\footnote{\autoref{eq:edot_generic} ignores the contribution of relative kinetic energy to the inflow of specific energy, which we will continue to assume throughout this paper. For the simulations explored here ($\mach=1/8-1/2$) this is always a small contribution, though it can become important in super-sonic mixing layers \citep{YangJi23}}.  Far above the interface, away from the turbulent motions, we expect this inflow to be relatively laminar. If $L$ is the length scale associated with the lateral extent of the layer then we can use $\Aint = L^2$ in \autoref{eq:edot_generic} to write down the velocity of the bulk inflow to the layer, far from the interface
\begin{equation}
    \label{eq:vbulk_def}
    \vbulk \equiv \frac{\gamma -1}{\gamma}\frac{\edot}{P L^2} \, ,
\end{equation}
where $\vbulk$ is understood to be measured in the frame co-moving with the layer.

An analog system to TRMLs, in which energy is injected into the fluid through chemical combustion processes, has been studied at length over the past century in the field of turbulent combustion \citep{kuo12,PoinsotBook}. In this literature, a key dimensionless quantity is the ratio of the eddy turnover time of integral-scale turbulent eddies, $\teddy(\Lint)$, to the cooling time, $\tcool$, the so-called Damk\"{o}hler number \citep{damkohler40}
\begin{equation}
    \label{eq:Da_def}
    \Da = \frac{\teddy(\Lint)}{\tcool} = \frac{\Lint}{\tcool \vtL}  \, .
\end{equation}
In \autoref{app:Da_def} we compare how this is measured in our simulations to that used in past works.

In the $\Da < 1$ limit, turbulent diffusion acts quickly in comparison to cooling and the turbulence is able to effectively ``smooth-out'' the interface on the largest scales available to it ($\Lint$). This is the ``well-stirred reactor'' or ``slow-cooling'' regime. In this limit, we can coarse-grain the fluid equations on scale $\Lint$, thereby smoothing out the interface area and considering it to be ``effectively'' laminar so that $\Aint \approx L^2$ (the actual interface area, on the smallest scales, should in fact be larger e.g. \citet{CPS91}). However, in this regime, the inflow velocity to the layer is determined by the balancing of the coarse-grained turbulent diffusion and cooling \citep{ZDPN69,TanOh21,Lancaster24a}
\begin{equation}
    \label{eq:vdiff_turb}
    \vdiff = \sqrt{\frac{\vtL \Lint}{\tcool}} \, .
\end{equation}
Using \autoref{eq:vdiff_turb} in \autoref{eq:edot_generic} as $v_{\rm in} = \vdiff$ and re-arranging we have
\begin{equation}
    \label{eq:edot_sc}
    \frac{\edot}{\vtL} = \left(\frac{\gamma}{\gamma - 1}\Pbar L^2 \right) \Da^{1/2}
\end{equation}
If the turbulence is considered fixed as we vary the cooling time over a series of experiments this would manifest as a $\edot \propto \Da^{1/2}$ dependence, as has been seen in many previous works \citep{FieldingFractal20,Tan21}. Past works generally agree on the scaling presented in \autoref{eq:edot_sc} and the interpretation presented above that leads to it.

In the $\Da > 1$ regime, the so called `corrugated flamelet' or `fast cooling' regime, turbulence on the largest scales evolves more slowly than the reaction can take place. The interface is then not able to be smoothed out by the turbulence but remains thin while the turbulence gives it multi-scale, `fractal' structure by advecting the thin surface with the flow. It is generally agreed based on numerical experiments in the astrophysical literature \citep{FieldingFractal20,Tan21,Das24,TMG25} and physical and numerical experiments in the turbulent combustion literature \citep{gulder91,zimont95,kuo12} that $\edot/\vtL \propto \Da^{1/4}$ in this regime. However, the interpretations that lead to this scaling are not consistent across different works. There are broadly two different mechanisms to explain the $\Da^{1/4}$ scaling, both working from \autoref{eq:edot_generic} and modifying it in different ways.

The first mechanism, first proposed by \citet{gulder91} and also preferred in the model of \citet{Tan21}, leaves the argument that applies in the $\Da < 1$ regime relatively unchanged. The only amendment is to replace $\tcool$ in \autoref{eq:vdiff_turb} with a new `effective' reaction time
\begin{equation}
    \label{eq:treact_effective}
    \ttcool = \sqrt{\tcool \teddy(\Lint)} \, ,
\end{equation}
the geometric mean of the large-scale eddy-turnover time and the reaction time. \citet{gulder91} arrive at this replacement\footnote{\autoref{eq:treact_effective} is only equivalent to the argument presented in \citet{gulder91} if the microphysical thermal and momentum diffusivity (kinematic viscosity) are assumed to be identical, i.e. a Prandtl number of 1.} by arguing that $\ttcool$ should be essentially given by the time it takes for a laminar flame, mediated by micro-physical diffusivity, to traverse the Taylor microscale, $\lT$, which can roughly be thought of as the small-scale end of the turbulent inertial range, or the typical spacing between dissipative vortices (which exist on the Kolmogorov length scale, $\lK$, on which turbulent diffusivity is balanced by viscosity).

\citet{Tan21} present a simpler argument which gives the same result by arguing that one can think of the large scale turbulence in the $\Da >1$ regime as an effective diffusivity, $D_{\rm turb} = \vtL \Lint$. A parcel of gas is then able to diffuse a distance
\begin{equation}
    \label{eq:lFt}
    \lFt = \sqrt{\vtL \Lint \tcool} \, 
\end{equation}
within a reaction time. They argue that over $\tcool$ only a fraction $f_{\rm cool}\approx\lFt/L$ of the layer is able to actually participate in the cooling process so that the more appropriate cooling time to assign to the whole volume is $\tcool/f_{\rm cool} = \tcool L/\lFt = \ttcool$ (as given in \autoref{eq:treact_effective}). As this is the relevant cooling time in their picture, they then replace $\tcool$ in \autoref{eq:vdiff_turb} with $\ttcool$, leading to the desired $\edot\propto\vtL\Da^{1/4}$ scaling. This argument picks out a special reaction time $\ttcool$ which should change as a function of $\Da$ (we return to this in \autoref{sec:discussion}).

The second mechanism to explain the scaling in the $\Da > 1$ regime is provided by \citet{FieldingFractal20}. Here the observation is made that for most realistic models of turbulence $\vtl$ increases sub-linearly as a function of scale, so that $\teddy(\ell)$ will become shorter as we move from the integral scale, $\Lint$, down. By this logic we will reach a scale, $\lcool$, at which 
\begin{equation}
    \label{eq:lcool_def}
    \teddy(\lcool) = \tcool\, ,
\end{equation}
or the scale on which $\Da = 1$ (considering turbulent eddies on that scale). At this point, we have reduced the problem to one which we have already solved since on this scale this portion of the surface behaves as though it is in the $\Da \lesssim 1$ regime. We then use \autoref{eq:edot_generic} with $\vdiff$ given by \autoref{eq:vdiff_turb} with $\vtL\Lint$ replaced with $v_t(\lcool) \lcool$, the only reasonable choice for turbulent diffusivity at this scale. The question then is what to use for $\Aint$ in \autoref{eq:edot_generic}. As we mentioned above, in the $\Da > 1$ regime the turbulence on scales larger than $\lcool$ acts to wrinkle the surface, giving it multi-scale, fractal structure. If we assume that the surface can be characterized by a single fractal dimension $D = d+2$ (with $0<d<1$ the `excess fractal dimension') between the scales $\Lint$ and $\lcool$ then the area of the surface on scale $\lcool$ is
\begin{equation}
    \Aint(\lcool) = L^2\left( \frac{\lcool}{\Lint}\right)^{-d} \, .
\end{equation}
If we additionally assume that the turbulence follows a power-law structure as a function of scale
\begin{equation}
    \label{eq:vsf_pl}
    \vtl = \vtL \left(\frac{\ell}{\Lint} \right)^p
\end{equation}
we can put these parts together into \autoref{eq:edot_generic} to arrive at
\begin{equation}
    \label{eq:edot_fractal}
    \frac{\edot}{\vtL} = \frac{\gamma}{\gamma - 1} P L^2 \Da^{\frac{d-p}{1-p}} \, .
\end{equation}
If we take $d=1/2$ as is measured in mixing layer simulations and $p=1/3$ as is appropriate for Kolmogorov turbulence then $(d-p)/(1-p) = 1/4$ and we have the desired scaling. We will see below that while the fractal dimension of $d=1/2$ does seem to apply over a broad range, it does not work everywhere. Additionally, the appeal to Kolmogorov scaling to describe multiphase, compressible turbulence is questionable and we will see that it does not apply in our simulations.

In a companion paper, which we will refer to throughout as \paperi, we presented a series of simulations of TRMLs and detailed our measurement of $\Aint$ and $\vdiff$ in the simulations and how they depend on numerical resolution. In particular, we demonstrated that a relation of the form of \autoref{eq:edot_generic} holds in our simulations with $\Aint$ and $\vdiff$ as measured on the grid scale. Resolution independence of total cooling is then explained by the countervailing resolution dependence of $\vdiff$ ($\propto \dx^{1/2}$) and $\Aint$ ($\propto \dx^{-1/2}$) in these simulations\footnote{We additionally showed how the full phase structure of the mixing layer (the amount of gas at intermediate temperatures and its thermal pressure) are not well resolved unless one is able to resolve $\lcool$.}. Having shown that $\edot$ is explained by the product of $\vdiff$ and $\Aint$, we now investigate what the scaling of these parameters with $\Da$ can elucidate about the origins of the $\edot/\vtL \propto \Da^{1/4}$ scaling behavior in the fast cooling regime.

In \autoref{sec:methods} we briefly review the details of our simulations suite. In \autoref{sec:results} we present the results of these simulations, specifically the scaling of $\vtL$, $\vdiff$ and $\Aint$ with $\Da$ and show that reduced cooling in the fast-cooling regime is due to decreased $\Aint$ in this regime due to the suppression of fractal structure on large scales. In \autoref{sec:discussion} we briefly discuss the consequences of these results in the context of past work and derive a new relation based on the results of \autoref{sec:results} which self-consistently predicts the emergence of the $\edot/\vtL \propto\Da^{1/4}$ scaling regime. Finally, we summarize our conclusions in \autoref{sec:conclusion}.

\begin{figure*}
    \centering
    \includegraphics[width=\textwidth]{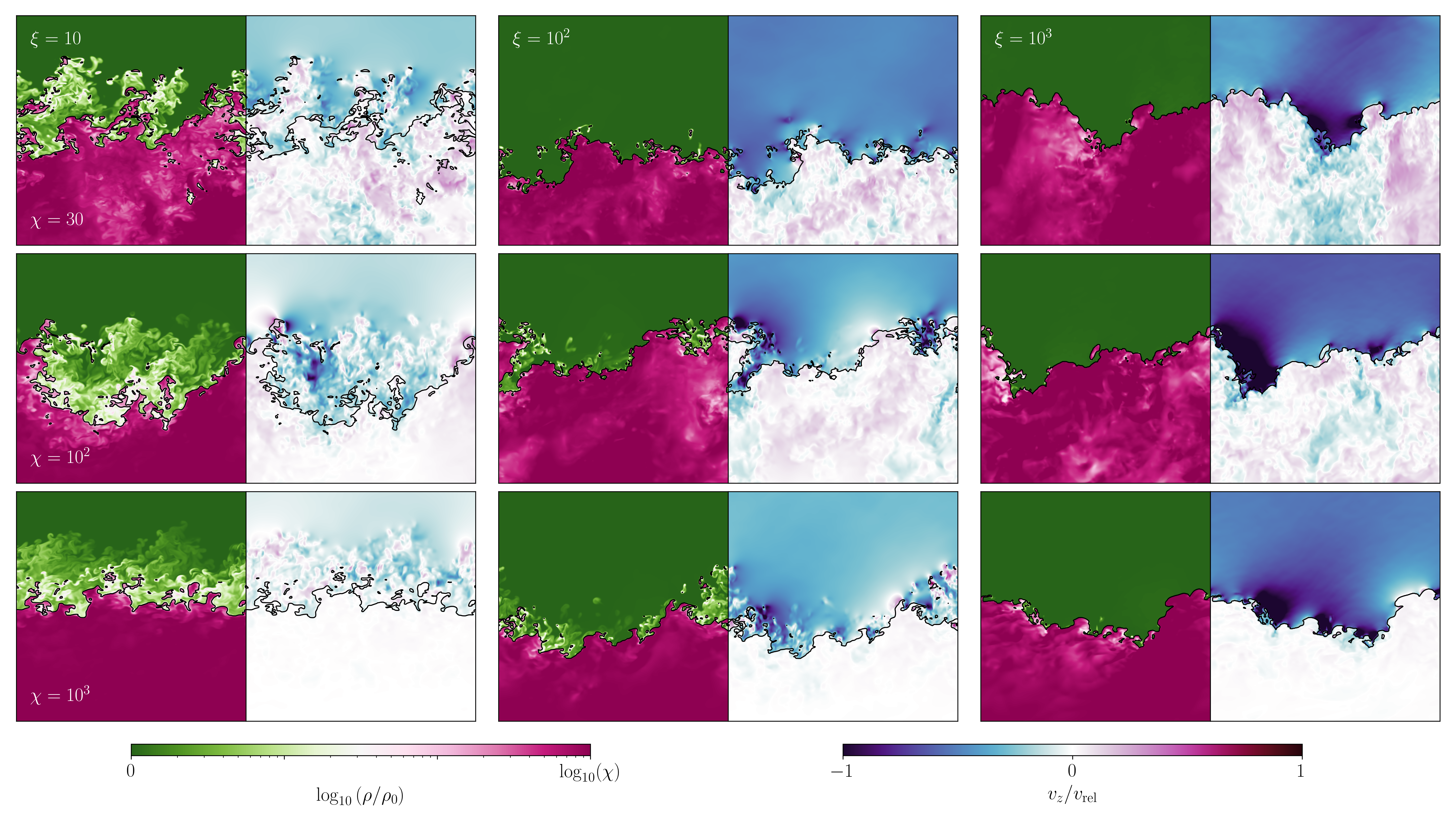}
    \caption{Slices through $y=0$ plane of a suite of simulations with $\nres = 512$, $\mach=1/2$ and varying $\chi$ and $\xi$ at $t/\tsh = 20$ showing density (pink/green) and vertical velocity (blue/pink). Rows show simulations with varying density contrast with $\chi = 30,\, 10^2,\, \&\, 10^3$ in the $1^{\rm st}$, $2^{\rm nd}$, and $3^{\rm rd}$ rows respectively. Columns show simulations with varying cooling strength with $\xi = 10,\, 10^2,\, \&\, 10^3$  in the $1^{\rm st}$ \& $2^{\rm nd}$, $3^{\rm rd}$ \& $4^{\rm th}$, and $5^{\rm th}$ \& $6^{\rm th}$ respectively. Each slice shows the full-extent of the simulations in $x$ but shows a window of length $\Lbox$ in the $z$-direction. Black lines in each plot indicate iso-temperature contours of $T=\Tpk$.}
    \label{fig:multi_panel}
\end{figure*}

\begin{figure}
    \centering
    \includegraphics{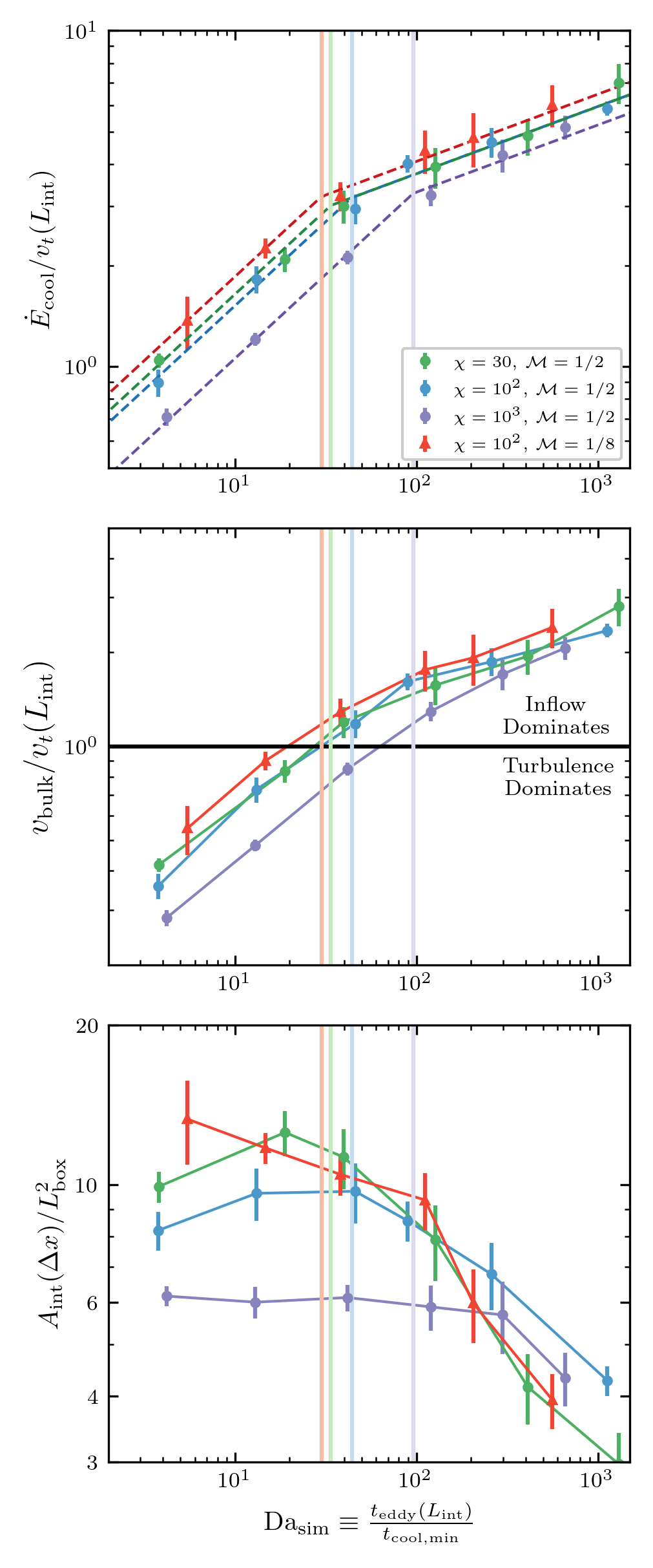}
    \caption{The behavior of cooling ($\edot/ \vtL$, top), the ratio of inflow velocity to the turbulent velocity ($\vbulk/\vtL$,middle panels), and interface area ($\Aint$, bottom),
    as a function of $\Da$ for the high resolution simulations ($\nres = 512$). Variations in  $\chi$ and $\mach$ are indicated by color in the top panel's legend. Scaling behavior of $\edot/\vtL$ with $\Da$ are shown as dashed lines in the top panel. The values at which scaling of $\edot$ with $\Da$ transitions are shown as vertical lines in all panels.}
    \label{fig:vdiff_AT}
\end{figure}

\begin{deluxetable}{cc}
\tablecaption{Parameters of simulation suite.\label{tab:sim_params}}
\tablewidth{0pt}
\tablehead{
\colhead{Parameter} & \colhead{Values}}
\startdata
$\chi$ & $30$, $100$, $1000$  \\
$\xi$ & $1$, $3$, $10$, $30$, $100$, $300$, $1000$  \\
$\mach$ & $1/8$, $1/2$ \\
$\nres$ & $64$, $128$, $256$, $512$
\enddata
\end{deluxetable}

\section{Simulation Description}
\label{sec:methods}

A more detailed description of our simulations is given in \paperi, but we review the essential aspects here for clarity. We perform a set of hydrodynamical simulations of TRMLs using the GPU-accelerated \ak code \citep{athenak}. The simulations are initialized within a box with side lengths $(L_x, L_y, L_z) = (\Lbox, \Lbox, 1.5\times\Lbox)$ at uniform resolution $\dx \equiv \Lbox/\nres$. The initial conditions consist of two states that are transitioned between at $z_{\rm trans}=\Lbox/2$ with the ``upper'' ($z>z_{\rm trans}$) state consisting of hot, low-density gas and the ``lower'' ($z<z_{\rm trans}$) state consisting of cold, high-density gas. These phases are initially in pressure equilibrium at pressure $\Pbar$ with the hot gas at density $\rho_{\rm hot} = \rhobar$ and the cold gas at density $\rho_{\rm cold} = \chi \rhobar$. We set temperature simply as $T = P/\rho$ so that $\Thot = \Pbar/\rhobar = \chi \Tcold$ (for pressure equilibrium). The adiabatic sound speed in the hot gas is $\cshot = \sqrt{\gamma\Pbar/\rhobar}$ with $\gamma = 5/3$. The two phases are set up in relative motion in the $x$-direction with $v_{x,{\rm hot}} = \vrel/2$ and $v_{x,{\rm cold}} = -\vrel/2$. Noise is introduced near the interface in order to excite the Kelvin-Helmholtz instability.

We employ a cooling function similar to that in \citet{CFB23} in which the two phases are thermally stable (heating and cooling balance) but cooling dominates at intermediate temperatures. The cooling function is a piece-wise power law with the volumetric cooling rate proportional to $\edotc \propto T^{-\beta(T)}$ with $(\blo,\, \bhi) = (-2,\, 3) \times 2/\log_{10}(\chi)$ for $\blo$ at $T<\Tpk$ and $\bhi$ at $T>\Tpk$. The $\chi$ dependence is included to assure that $\edotc(\Thot)$ and $\edotc(\Tcold)$ are the same across simulations with varying $\chi$. The cooling rate at the peak cooling temperature $\Tpk \equiv \left(\Tcold^2\Thot\right)^{1/3}$ is determined by the minimum cooling time, $\tcoolmin$, which is itself set by a choice of the parameter \citep{FieldingFractal20}
\begin{equation}
    \label{eq:xi_def}
    \xi \equiv \frac{\tsh}{\tcoolmin}
\end{equation}
where $\tsh \equiv \Lbox/\vrel$ is the shear-time of the layer. This parameter is very similar to $\Da$ and we explore their relationship in \autoref{app:Da_def}. In \autoref{app:tcool} we briefly explore the impact of the variation in the shape of the cooling function on integrated quantities ($\edot$). While we expect this shape to have an impact on the phase structure of the gas (explored in more depth in \paperi) we do not expect it to impact any of the main conclusions of this work, though this would be important to check in future studies.

We consider $\Pbar$ and $\rhobar$ to be fixed across the simulations (at unity in code units) and therefore the parameters that determine a simulation are the Mach number of the shear flow $\mach \equiv \vrel/\cshot$ (which sets $\vrel$), the density contrast $\chi \equiv\rho_{\rm cold}/\rho_{\rm hot}$ (which sets $\rho_{\rm cold}$), and $\xi$ (which sets $\tcoolmin$). We run simulations at $\mach = 1/2$, $\xi = 1,\, 3,\, 10,\, 30,\, 100,\, 300,\,\&\, 1000$, and $\chi = 30,\, 10^2,\, \&\,10^3$, though we do not discuss the $\xi = 1$ simulations here. We additionally run simulations at $\mach = 1/8$ for the $\chi = 10^2$ runs. Each parameter choice above is run at resolutions of $\nres = 64,\, 128,\, 256,\, \&\, 512$. The simulations are run for $30\,\tsh$ and all analysis is performed on $t> 10\, \tsh$ in order to assure the layer has reached an approximate steady-state. In \paperi\ we primarily explored the resolution dependence of certain properties of TRMLs. Here we will focus on how scaling behavior in TRMLs vary as a function of $\Da$ and $\chi$ and will only explore the resolution dependence when necessary. In \autoref{fig:multi_panel} we show slices in $\rho$ and $v_z$ through the $y=0$ plane of several of our high-resolution simulations with varying $\chi$ and $\xi$.

We measure the area of the interface between the phases, $\Aint$, by using the marching-cubes algorithm to identify iso-temperature surfaces at $T= \Tpk$ \citep{lorensen87,Chernyaev95,Lewiner03_marching_cubes}. In order to get a sense of the fractal structure of the mixing layers, we also use the \citet{scikit-image} implementation\footnote{In order to make sure that this calculation properly takes into account that the simulations are periodic in $x$ and $y$ we must supplement the simulation domains by a copy of one layer of cells from the left end of the simulation to the right end in both $x$ and $y$. If one does not account for this it can lead to significant bias for measurements at large scales $\ell$.} to calculate $\Aint(\ell)$ in all snapshots output from our simulation for scales $\ell = \dx - \Lbox/4$ in increments of factors of 2 (i.e. $\ell = \dx,\, 2\dx,\, 4\dx,...,\Lbox/8,\, \Lbox/4$). With these measurements we can measure the excess fractal dimension of the surface as
\begin{equation}
    \label{eq:frac_measure}
    d = - \frac{d\log \Aint(\ell)}{d\log \ell} \, ,
\end{equation}
which is implicitly itself a function of scale, $\ell$. In general we will measure the interface area at the resolution scale, $\Aint(\dx)$.

Details of the measurement of turbulent structure functions is given in \paperi\ Section 4.3. Briefly, we measure the 2nd-order structure functions of the three components of the velocity field separately, for $t>10\tsh$, and restricting to the mixed gas. We define $\vtl$ as it is used in \autoref{sec:results} as $\vtl = \sqrt{3}{\rm SF}_2(v_y)$ where, similarly to \autoref{eq:vsf2_def}, ${\rm SF}_2(f)$ is defined as the square-root of the traditional 2nd order structure function so that ${\rm SF}_2(f)$ has the same dimensions as $f$.

The velocity at which gas is carried through the surface by numerical diffusion is measured by considering the difference in the velocity into the layer just above and just below the interface (\paperi\ provides further explanation of this measurement). We demonstrate that this measurement is accurate by showing that, in combination with $\Aint$, it correctly predicts the total cooling rate in the layer
\begin{equation}
    \label{eq:edot_measure}
    \edot \equiv \int_V \left(\edotc - \edoth \right) dV \, .
\end{equation}

\section{Results}
\label{sec:results}

\subsection{$\edot(\Da)$ is Determined by $\Aint(\Da)$}
\label{subsec:res1}

In \autoref{fig:vdiff_AT} we show $\edot/\vtL$ (top panel), the ratio of bulk inflow velocity to turbulent velocity $\vbulk/\vtL$ (middle panel), and $\Aint(\dx)$ (bottom panel) as a function of $\Da$ and $\chi$ in our highest resolution simulations. We measure the integral scale of the turbulence in our simulations by the scale that maximizes $\vtl$ (details on measurement of $\vtl$ in \autoref{sec:methods} and Section 4.3 of \paperi). We then measure $\Da$ in our simulations using the eddy turnover-time on this integral scale as:
\begin{equation}
    \label{eq:Da_measure}
    \Dasim = \frac{\teddy(\Lint)}{\tcoolmin} \, .
\end{equation}
This will be used to indicate the value as measured in the simulations throughout the rest of the work. In \autoref{app:Da_def}, we discuss other methods of measuring $\Dasim$ and its relationship to $\xi$. In \autoref{app:tcool} we discuss the use of $\tcoolmin$ as the relevant cooling time.

$\edot/\vtL$ is fit as a function of $\Da$ for each simulation suite with equal values of $\chi$ and $\mach$ where we assume a piecewise power-law with exponent $1/2$ below a transition and exponent $1/4$ above a transition. We fit for the amplitude of the curve and the transition point between the two regimes, these fits are shown as colored dashed lines in the top-panel of \autoref{fig:vdiff_AT}. The transition points are marked as vertical lines in all panels of \autoref{fig:vdiff_AT}.

\begin{figure*}
    \centering
    \includegraphics[width=\textwidth]{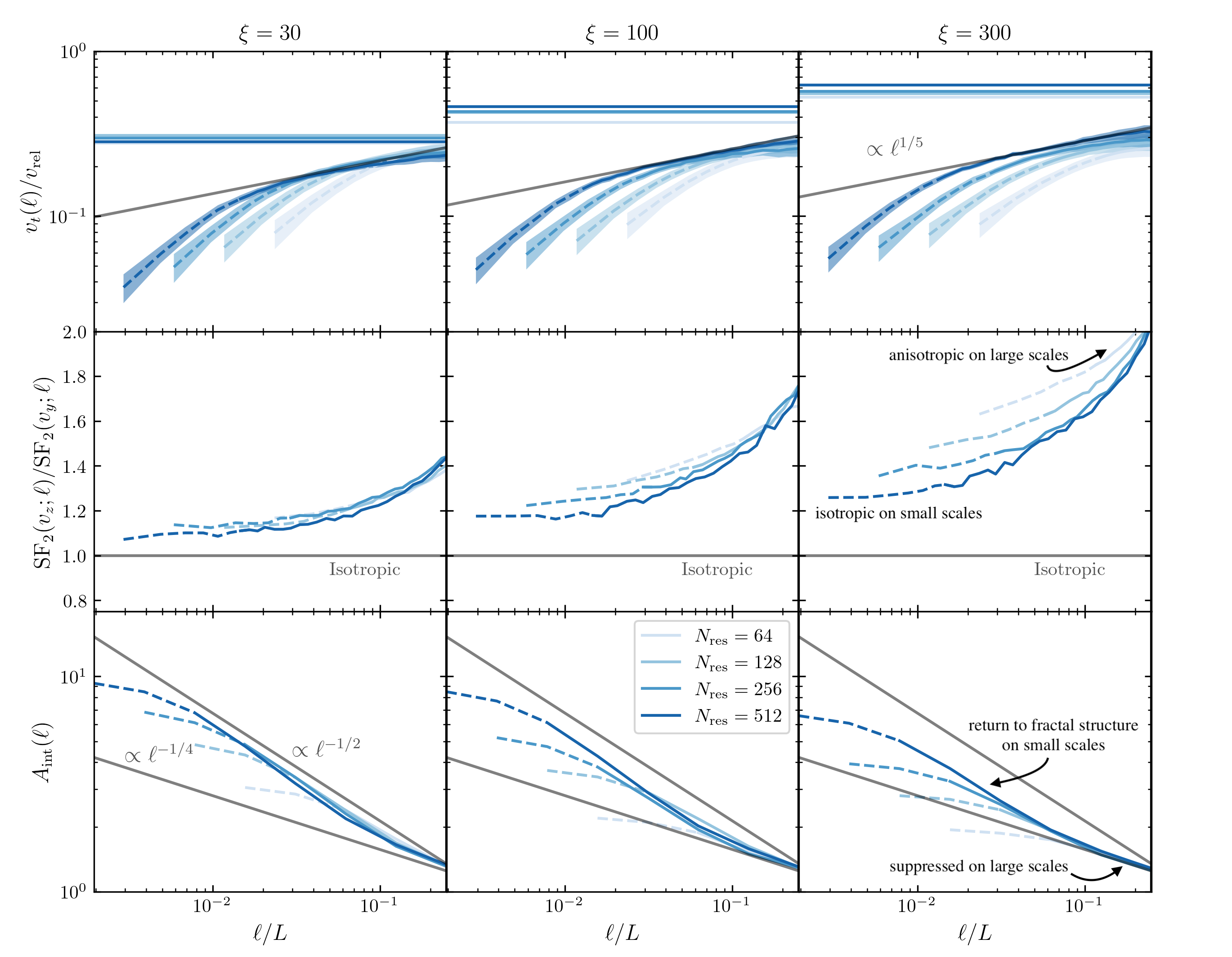}
    \caption{The velocity SFs and interface surface areas, as a function of scale in the $\mach = 1/2$, $\chi=10^2$ simulations that are closest to the transition in scaling behavior from $\edot \propto\Da^{1/2}$ to $\edot\propto \Da^{1/4}$, specifically $\xi = 30,\, 100,\, \&\, 300$ (left to right). \textit{Top panels}: The turbulent SFs with scale, shaded regions correspond to $1\sigma$ deviations over time, darker lines correspond to higher resolution. Horizontal lines indicate the inflow velocity from the top boundary of the box, $\vbulk$, defined in \autoref{eq:vbulk_def}. Gray lines in each panel indicate a $\vtl \propto \ell^{1/5}$ power-law dependence on scale. \textit{Middle panels}: The ratio of 2nd-order structure functions of $v_z$ to that of $v_y$. The horizontal gray line indicates isotropy. \textit{Bottom panels}: The interface surface area as a function of scale, $\Aint(\ell)$ (measurement described in \paperi). Gray lines indicate the expected scale dependence for fractal surfaces with $d= 1/4$ (lower line) and $1/2$ (upper line). Scales $\ell < 8\dx$ ($4\dx$) are indicated by dashed lines in the top and middle (bottom) panels. This is to indicate roughly the end of the turbulent inertial range (top and middle panels) and the end of self-similar fractal structure (bottom panels).}
    \label{fig:vsf_Asup}
\end{figure*}

Some points of note from \autoref{fig:vdiff_AT} are
\begin{itemize}
    \item[1.] Top panel: Despite all simulations being in the $\Da \gtrsim 1$ ``fast-cooling'' regime $\edot/\vtL$ transitions in scaling behavior from $\edot/\vtL\propto \Da^{1/2}$ to $\edot/\vtL\propto \Da^{1/4}$ in the all simulations.

    \item[2.] Middle panel: The transition in scaling behavior that is fit to the points in the top panel corresponds approximately to the point at which the bulk inflow velocities, $\vbulk$ (\autoref{eq:vbulk_def}), become greater than the integral-scale turbulent velocities, $\vtL$.

    \item[3.] Bottom panel: The transition in scaling behavior of $\edot/\vtL$ also corresponds to a transition in the behavior of $\Aint(\dx)$ with $\Da$ from roughly constant at low $\Da$ to being progressively suppressed above the transition.
\end{itemize}
As we discuss below around \autoref{fig:vsf_Asup}, the level of turbulence, $\vtL$, is also a very weak function of $\Da$ and $\vbulk$ becomes comparable to $\vrel$ in the high $\Da$ limit.

In \paperi\ we showed that, in analogy with \autoref{eq:edot_generic},
\begin{equation}
    \label{eq:edot_paper1}
    \edot = \frac{\gamma}{\gamma -1}\Pbar \vdiff \Aint(\dx)
\end{equation}
with $\vdiff \propto \dx^{1/2}\Da^{1/2}$ caused by numerical diffusion. Considering this consistent scaling of $\vdiff$ with $\Da$ across the simulation suite together with \autoref{eq:edot_paper1}, the approximate independence\footnote{While we do not show it explicitly here our simulations indicate a very mild scaling $\vtL \propto \Da^{1/10}$ in the $\Da\gg 1$ regime, consistent with \citet{Tan21}.} of $\vtL$ on $\Da$, and the change in $\Aint(\dx)$ behavior corresponding to the change in $\edot/\vtL$ scaling, it seems abundantly clear that the change in $\edot/\vtL$ scaling is caused by the suppression of $\Aint(\dx)$ at high $\Da$.

\subsection{What Suppresses $\Aint(\Da)$?}
\label{subsec:res2}


What are the driving factors behind the value of $\Aint$? As noted above, we see from the bottom panel of \autoref{fig:vdiff_AT} that $\Aint$ has a roughly constant value below the transition and that this value is smaller for larger $\chi$. Given that each mixing layer has roughly the same level of turbulence, $\vtL$, this lower $\Aint$ at higher $\chi$ is likely caused by the inertia of the cold gas: it is harder to ``mix-up'' (and therefore fold, and enhance the interface area of) gas with a larger inertia. This effect manifests not only as lower $\Aint$ but a smaller vertical extent of the mixing layer, as is visually apparent in \autoref{fig:multi_panel}.

As we already noted in \autoref{subsec:res1}, in the high $\Da$ regime the bulk inflow velocity to the layer becomes comparable to the integral scale turbulent velocity, $\vtL$. We might expect this to significantly impact the turbulent folding of the interface once the ram pressure of the inflow ($\rho_{\rm hot} \vbulk^2$) becomes comparable to or greater than the turbulent pressure ($\approx\rho_{\rm cold}\vtL^2$, assuming the dynamical pressure is dominated by the cold phase), so that the suppression of interface growth occurs when $\vbulk \gtrsim \sqrt{\chi} \vtL$.

The above argument is somewhat maximal in that it compares the ram-pressure of the low density inflow to the turbulent pressure of the cold, high density gas. In reality, turbulent motions in the hot gas should be significantly disrupted once the inflow's ram-pressure is comparable to the turbulent pressure \textit{in the hot phase} $\rho_{\rm hot}\vbulk^2 \gtrsim \rho_{\rm hot} \vtL^2$. Since the flow is subsonic, disruption of turbulence in the hot gas likely suppresses turbulence in the cold gas, as cold-phase motion that is strongly countervailing to motion in the hot-phase would result in compression and pressure gradients that would counter this motion. In this interpretation, we should have a suppression of structure when
\begin{equation}
    \label{eq:suppression_estimate}
    \vbulk \gtrsim \vtL \, ,
\end{equation}
which is exactly what is observed in \autoref{fig:vdiff_AT}.

In \autoref{fig:vsf_Asup} we further illustrate the way in which the inflow suppresses $\Aint$ in the $\chi=10^2$ simulation. In the top panels we show the turbulent SFs, $\vtl$, in comparison to $\vbulk$ (horizontal lines) over the range of simulations where the change in $\Aint(\Da)$ scaling behavior occurs ($\Da \approx \xi = 30-300$). We include the full set of simulation resolutions here in order to illustrate what part of $\vtl$ is well-resolved. In particular, we indicate all scales $\ell \leq 8\,\dx$ with dashed-lines as these seem to roughly correspond to the range where motions are damped due to numerical viscosity. Consistent with the resolution independence of $\edot$ discussed in \paperi, $\vbulk$ is also relatively independent of resolution across these simulations.

In the middle panels of \autoref{fig:vsf_Asup} we show the ratio of the 2nd-order structure functions in the $z$- and $y$-components of the velocity field. As discussed in \paperi,  $v_y$ is considered the independent tracer of turbulence as it is not associated with the shear or inflow directions. Therefore, the ratio we show in the middle panels of \autoref{fig:vsf_Asup} is a tracer of the degree of anisotropy in the turbulence as a function of scale. It is clear that all simulations shown in \autoref{fig:vsf_Asup} are anisotropic on large scales, indicating the importance of the inflow, while they are closer to isotropy on small scales, especially at low $\xi$.

Finally, in the bottom panels of \autoref{fig:vsf_Asup} we show how the interface surface area, $\Aint$, varies as you measure it on different scales (details in \autoref{sec:methods}). We indicate the expected scaling for fractals with $d=1/4$ and $d=1/2$ as gray lines. We see that, as we move to higher $\xi$ simulations, the fractal dimension of the surface is suppressed on the largest scales, with $\Aint \propto \ell^{-1/4}$, before returning to a $\Aint \propto \ell^{-1/2}$ scaling on intermediate scales and finally flattening out due to numerical viscosity on the smallest scales (as indicated by the dashed lines). This suppression at large scales corresponds exactly to where the velocity field begins to be strongly anisotropic ${\rm SF}_2(v_z)/{\rm SF}_2(v_y) \gtrsim 1.5$, indicating that it is the dynamical impact of the inflow that is suppressing $\Aint(\dx)$.

\begin{figure}
    \centering
    \includegraphics{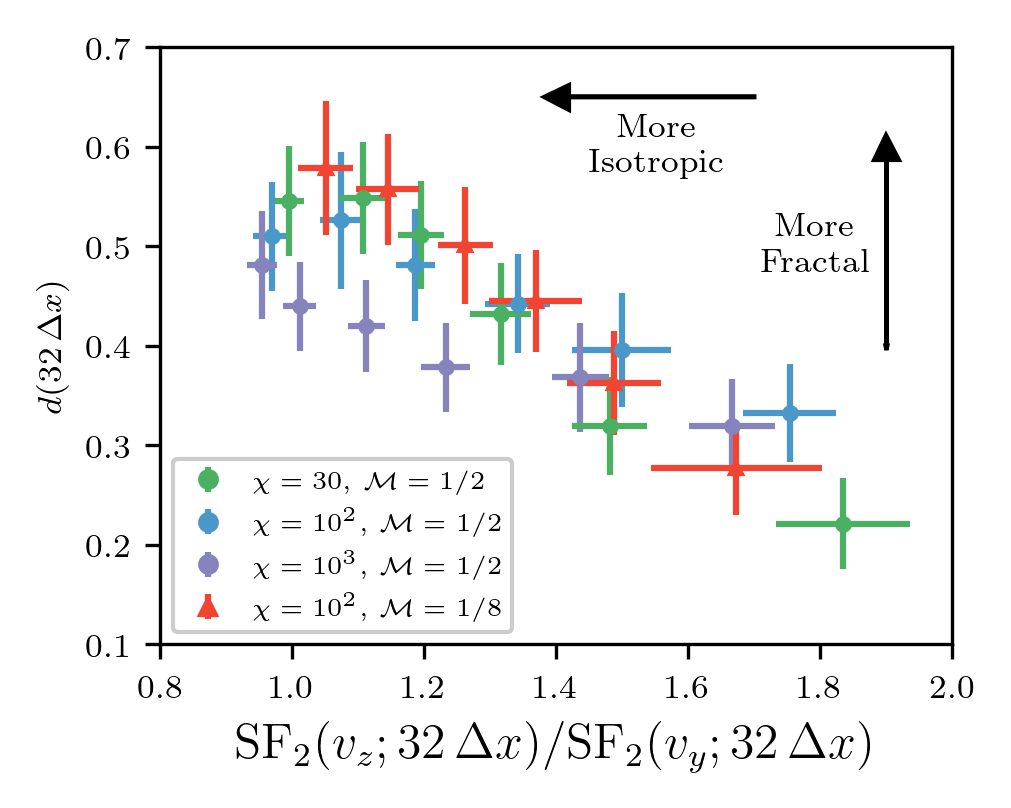}
    \caption{We show the excess fractal dimension of the interface, $d$ (\autoref{eq:frac_measure}), versus the ratio of structure functions (a measure of anisotropy) both measured on resolved scales ($32\,\dx$) in our highest resolution simulations ($\nres=512$). Errors are derived from standard deviations in each quantity measured over all snapshots with $t/\tsh>10$. We see that more isotropic velocity structure is very well correlated with more fractal structure in the interface (higher $d$).}
    \label{fig:frac_aniso}
\end{figure}

To illustrate this point further, in \autoref{fig:frac_aniso} we directly show the fractal dimension of the interface, $d$, measured using finite differences in log-space (\autoref{eq:frac_measure}) versus the anisotropy of the turbulence, both measured on resolved scales ($\ell = 32\, \dx$). It is clear that more isotropy leads to more fractal structure in the interface. The return to more volume-filling fractal scaling on the smallest scales is then natural given that the turbulence becomes more isotropic on these scales. This can be intuitively explained by the fact that the smallest scale structures are somewhat shielded from the inflow suppression, as they lie deeper within the turbulent structure of the interface (see \autoref{fig:multi_panel}). Indeed the return to a fractal-like scaling on the smallest scales is needed to explain the consistent $\Aint(\dx) \propto \nres^{1/2}$ scaling that we still observe in the highest $\xi$ simulations (see \paperi). However, the resolution-dependent $\chi=10$, $\xi = 10^3$ simulations presented in Appendix A of \paperi\ do not return to near-isotropy nor self-similar fractal scaling on small scales, explaining the resolution independence of $\Aint(\dx)$ and therefore resolution \textit{dependence} of $\edot$ exhibited there. It is also reassuring to see that the trend of decreased $d$, as well as the range of scales over which it occurs, is a consistent function of resolution in \autoref{fig:vsf_Asup}.

In the top panels of \autoref{fig:vsf_Asup} we also include lines indicating power-law dependencies of the turbulent velocity on scale, $\vtl \propto \ell^{1/5}$ as gray lines. We refer to the region well fit by this scaling as the ``inertial range'' of the turbulence, before it has a much steeper dependence due to numerical viscosity at smaller scales. This same scaling approximately holds for the ``inertial range'' of all the simulations explored in this work. These simulations appear not to follow Kolmogorov-like $\vtl \propto\ell^{1/3}$ scaling. This is not particularly surprising given the highly compressible and multiphase nature of the turbulence. Additionally, at high $\Da$ the ram-pressure of the inflow begins to have an important dynamical effect on the turbulence, as we argue above. This is surely expected to limit the actual inertial range (which is truly the range of scales over which the only important forces on the fluid are the inertial forces from the non-linear advection of eddies \citep{Frisch95}) by making the compressive Reynolds stresses dynamically important over a larger range of scales.

\section{Discussion}
\label{sec:discussion}

\subsection{Past Works}
\label{subsec:past_work}

In \autoref{sec:results} we demonstrated that the origin of the change in scaling behavior of the cooling as a function of $\Da$ is a suppression of the surface area of the mixing interface caused by the ram-pressure of the inflow disrupting the turbulent folding of the surface. Furthermore, from \autoref{fig:vsf_Asup}, it seems that the nature of this suppression in area, which occurs mostly on large scales, is well-resolved in the simulations.


It is fair to interrogate if this finding is congruent with the theoretical models of \citet{Tan21} and \citet{FieldingFractal20} discussed in \autoref{sec:intro}. In \autoref{subsec:res2} we have directly measured key parameters of the \citet{FieldingFractal20} model: the fractal dimension of the interface, $d$, and the power-law scaling of the turbulent structure function, $p$. We find that the $p = 1/3$, Kolmogorov-like scaling that is assumed in the \citet{FieldingFractal20} \textit{does not} apply in this regime\footnote{There is of course the possibility that our inferred $p$ is affected by our finite resolution \citep{Kritsuk07}.}. This is not necessarily surprising given that the flow is multi-phase and compressible. Further, while the excess fractal dimension of $d=1/2$ that is assumed in \citet{FieldingFractal20} does apply broadly in our simulations, it is precisely in the high $\Da$ limit that it ceases to apply at all scales (\autoref{fig:vsf_Asup}). While the exact argument of \citet{FieldingFractal20} does not seem to hold, we will see in \autoref{subsec:new_interp} that the general approach of balancing the inflow of hot gas against the fractal structure and turbulent diffusivity can still lead to a prediction consistent with the data presented here.

It is harder to confront the phenomenological argument of \citet{Tan21} with our available measurements. As discussed in \autoref{sec:intro}, this argument picks out a specific time-scale for parcels of gas that are cooling in the mixing layer: $\ttcool$ (\autoref{eq:treact_effective}). The best test of this model would be to track the cooling of Lagrangian tracer particles in the flow and see if their cooling histories, as an ensemble, had a characteristic time-scale for evolution of $\ttcool$. However, given that the argument of \citet{Tan21} makes no reference to the suppression of $\Aint$ in the high $\Da$ regime, it is not immediately obvious that their picture can be reconciled with the evidence presented here.

Finally, the work of \citet{Sharma25} finds (see Figure 5 of their supplementary material) that when increasing $\xi$ by increasing the transverse dimension of the box, $\Lbox$, (and thereby increasing $\tsh$) one does not find the $\edot\propto \xi^{1/4}$ scaling behavior expected (if $\xi \approx \Da$) but instead $\edot$ independent of $\xi$. It is our suspicion that this occurs exactly because of the effect that we discuss here: the suppression of growth in the turbulent structure due to the ram-pressure of the inflowing gas. In particular, in this regime, the vertical extent of the layer is suppressed and therefore the outer scale of the turbulence ceases to track the size of the box and instead tracks the thickness of the layer. \citet{Sharma25} also explain this result by appealing to the suppressed vertical extent of the layer but extend this to infer that the $\edot\propto \xi^{1/4}$ scaling regime does not exist. We would suggest that if the turbulent structure was measured in these simulations and $\Da$ as defined by \autoref{eq:Da_def} were used in place of $\xi$ that one would find that $\Da$ did not change across this range of experiments that increase $\Lbox$, consistent with the lack of changes in integrated cooling observed in that work. Though we do not explore the $\Da <1$ regime here, one may imagine that this same point may become important in this regime as the layer is expected to thicken vertically in this regime. In this case it is important to make sure that one is using a consistent definition of $\Da$.

\begin{figure}
    \centering
    \includegraphics{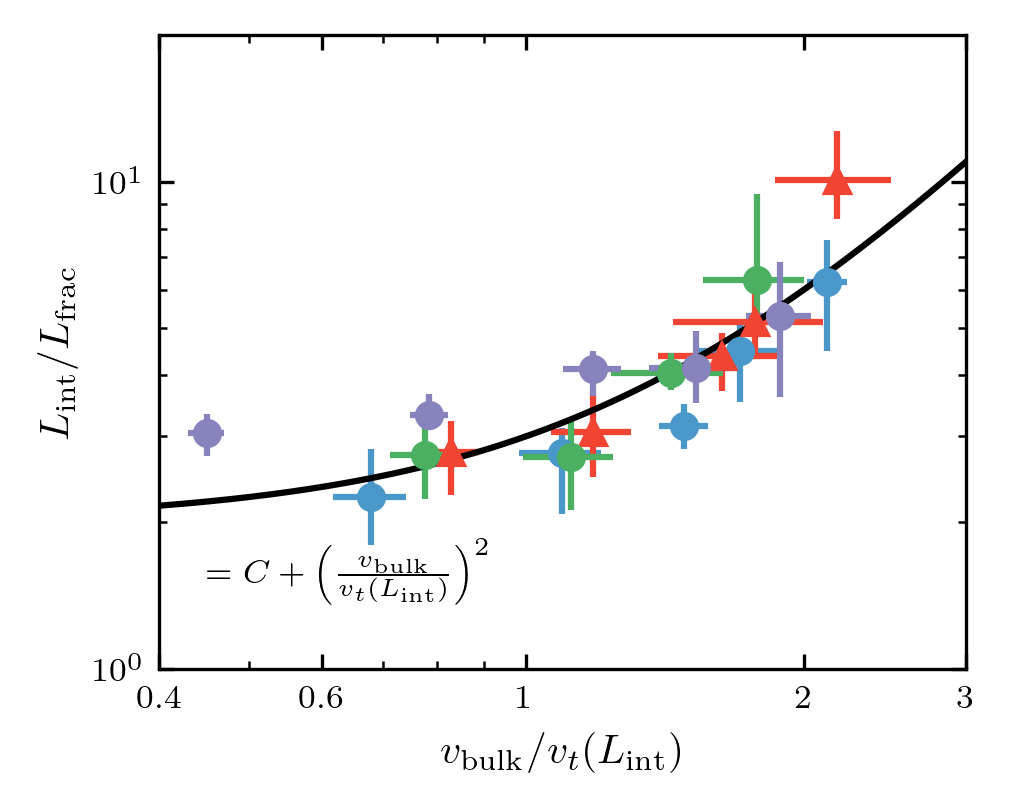}
    \caption{We show the ratio of the integral scale of the turbulence, $\Lint$, to the scale at which fractal structure sets in, $\Lf$, as a function of the bulk inflow velocity, $\vbulk$,  relative to the turbulent velocity, $\vtL$. Measurements are detailed in the text. We see that in the $\vbulk > \vtL$ regime $\Lf$ begins to become increasingly smaller in comparison to $\Lint$, indicating the suppression of fractal structure on large scales. The measured scaling of $\Lint/\Lf$ with $\vbulk/\vtL$ is consistent with the ram-pressure ansatz made in \autoref{eq:Lfrac_scaling} (black line) that shows a constant $C=2$ for $\vbulk < \vtL$ and the quadratic scaling above this value.}
    \label{fig:Lfrac_vbulk}
\end{figure}

\subsection{A New Interpretation}
\label{subsec:new_interp}

We now appeal to the phenomena presented in this work, namely the suppression of the fractal surface area, to explain the $\edot/\vtL \propto \Da^{1/4}$. In analogy with \autoref{eq:edot_generic}, we imagine a coarse-graining on scale $\ell$ of the dynamics across the layer and write the total cooling in the layer as a function of scale, $\ell$, as
\begin{equation*}
    \edot(\ell) = \frac{\gamma}{\gamma - 1}\Pbar \vdiff(\ell) \Aint(\ell) \, ,
\end{equation*}
where $\vdiff(\ell)$ is some \textit{effective} diffusivity on scale $\ell$, and $\Aint(\ell)$ is the interface area measured on scale $\ell$. Here we are imagining a ``coarse-graining'' of the surface on scale $\ell$, where the properties of the interface are smoothed on this scale and we imagine measuring the interface area and diffusive velocities similarly to how it is done in the simulations. We will assume that the effective diffusivity is mediated by turbulent motions on scale $\ell$ so that, in analogy with \autoref{eq:vdiff_turb}, $\vdiff(\ell) = \sqrt{\vtl\ell/\tcoolmin}$. However, differing from the argument of \citet{FieldingFractal20}, we will now assume that the fractal scaling of the interface only applies below a scale $\Lf$, so that
\begin{equation}
    \label{eq:Aint_supp}
    \Aint(\ell) = L^2 \left( \frac{\ell}{\Lf}\right)^{-d} \, ,
\end{equation}
where $L$ is still the large-scale dimension of the interface (equivalent to $\Lbox$ in our simulations) and we have assumed a scale hierarchy $\lambda_{\rm F}\ll \ell < \Lf < \Lint < L$, where $\lambda_{\rm F}$ is the true Field-length corresponding to the balance of micro-physical thermal conduction and cooling \citep{Field65}. This lower bound is instilled here to guarantee that the \textit{effective} turbulent diffusivity is the principal means of energy transport across the layer.

Combining the coarse-grained energy flux relation and \autoref{eq:Aint_supp} with a power-law scaling of the turbulent structure (\autoref{eq:vsf_pl}) we arrive at
\begin{equation}
    \label{eq:scaling1}
    \frac{\edot(\ell)}{\frac{\gamma}{\gamma-1}\Pbar L^2 \vtL} = \Da^{1/2}
    \left( \frac{\ell}{L}\right)^{\frac{1+p}{2} - d}
    \left( \frac{L}{\Lf}\right)^{-d} \, ,
\end{equation}
where $p$ is the power-law scaling of the turbulent velocity with $\ell$.

Let us imagine running a series of experiments, like those presented in the work, where we vary the cooling time, $\tcoolmin$, and hence $\Da$, while keeping the integral scale turbulent velocities, $\vtL$, fixed\footnote{While $\vtL$ increases mildly as one moves further in to the fast-cooling regime, $\Da\gg 1$, this dependence is very mild $\vtL\propto \Da^{1/10}$, (see also \citet{Tan21} Figure 12).}. Focusing now only on terms dependent on $\Da$, we can ignore the $(\ell/L)$ term.

As we showed in \autoref{subsec:res2}, at high $\Da$ the fractal structure becomes suppressed by the ram-pressure of the inflow disrupting the turbulence in the hot phase. It is natural to expect that the scale on which fractal structure is realized would then become suppressed in proportion to the ratio of the ram-pressure of the inflow to the turbulent pressure in the hot phase. So that, when $\vbulk > \vtL$ we have:
\begin{equation}
    \label{eq:Lfrac_scaling}
    \frac{L}{\Lf} \propto \frac{\vbulk^2}{\vtL^2} \, .
\end{equation}
We show the relationship between these two quantities\footnote{Ideally we would measure $\Lf$ in our simulations as the scale at which the fractal dimension of the interface, $d$ (\autoref{eq:frac_measure}), went from zero on large-scales to a non-zero value. In practice there is some level of structure on all scales and the transition is not as immediate as the relation shown in \autoref{eq:Aint_supp} implies. Therefore we measure $\Lf$ in the simulations as the scale at which $d(\Lf) > 1/3$ for a given snapshot and, in keeping with other measurements, take $\Lf$ as the median value over all snapshots with $t/\tsh > 10$.} in \autoref{fig:Lfrac_vbulk}. Indeed we see that when $\vbulk \gtrsim \vtL$ the scale at which fractal structure begins to apply, $\Lf$, becomes suppressed relative to the outer scale of the turbulence in a proportion consistent with \autoref{eq:Lfrac_scaling}.

Given the definition of $\vbulk$ in \autoref{eq:vbulk_def}, the right-hand side above is exactly the square of the left-hand side of \autoref{eq:scaling1}. Using \autoref{eq:Lfrac_scaling} in \autoref{eq:scaling1}, and ignoring the $\ell/L$ dependence, we have
\begin{equation}
    \label{eq:scaling2}
    \frac{\edot(\ell)}{\vtL} 
    \propto \Da^{\frac{1}{2(1 + 2d)}} \, ,
\end{equation}
where we have additionally dropped the constant parameters $\gamma$, $L$, and $\Pbar$. Taking $d=1/2$ in \autoref{eq:scaling2}, as we expect to apply at scales below $\Lf$ which is implicit in our derivation above, recovers $\edot/\vtL \propto \Da^{1/4}$, as has been observed in \autoref{sec:results} and several past works \citep{Tan21,FieldingFractal20}.

Based on the results presented in the top panel of \autoref{fig:vdiff_AT} it seems plausible that many of the simulations follow a somewhat shallower relation than $\edot/\vtL \propto \Da^{1/4}$. Indeed, allowing for freedom in the high-$\Da$ value of the power-law index to our piecewise power-law fits in the top-panel of \autoref{fig:vdiff_AT} results in exponents which are closer to $1/5$ ($0.17-0.22)$, while maintaining the order and approximate value of the transitions in behavior between different simulation suites. This freedom of interpretation in the scaling naturally calls in to question the validity of applying the $\Da^{1/4}$ scaling far outside of the regime in which it has been tested with TRML simulations (e.g. $\Da \gg 10^3,\, \chi \gg 10^3$ \citet{Lancaster21a},\citet{MLD25}).

We conclude with a note on the $\ell/L$ term in \autoref{eq:scaling1}. Within this coarse-grained model, requiring $\edot(\ell)$ to be independent of the arbitrary coarse-graining scale $\ell$ implies a consistency condition between the turbulent structure and the interface geometry, $d=(1+p)/2$. We view this as a model requirement rather than a direct empirical result. It is interesting that a similar relation is suggestively reminiscent of Yaglom's Law \citep{Yaglom49}\footnote{Drawing this relation additionally requires the relationship between the H\"{o}lder exponent of the temperature field, $\beta$ ($\delta T \propto \ell^{\beta}$), and the fractal dimension of its iso-sets $D = 3 - \beta$ \citep{FedererGMT,DeLellis23}, with $D = 2+ d$.}, which relates velocity statistics to those of an advected scalar field \citep{Warhaft00,MoninYaglomBook}; we take that correspondence as motivation rather than proof in the present compressible, multiphase flow. On the smallest resolved scales of the low-$\Da$ simulations, where the flow is most nearly isotropic, we measure $d=0.5-0.6$, with the upper end of this range being broadly consistent with $p=1/5$ (\autoref{fig:vsf_Asup}) and the above consistency relation. This correspondence is only noted as it is far too interesting not to note, though it is inconsequential to the rest of our argument.

\section{Conclusion}
\label{sec:conclusion}

Our main conclusions are as follows:
\begin{itemize}
    \item[1.] We demonstrate (\autoref{fig:vdiff_AT}) that the change in behavior of $\edot/\vtL$ with $\Da$ occurs near the point where the velocity of gas inflowing to the layer (to balance cooling occurring in the layer), $\vbulk$, becomes greater than the integral-scale turbulent velocity in the layer, $\vtL$.

    \item[2.] At this same transition, the area of the interface between hot and cold gas, $\Aint$, becomes suppressed in comparison to its low-$\Da$ (roughly constant) value (bottom panel of \autoref{fig:vdiff_AT}).

    \item[3.] The suppression of $\Aint$ tracks, as a function of physical scale, the anisotropy of turbulent structure in the velocity field (\autoref{fig:vsf_Asup} and \autoref{fig:frac_aniso}), caused by the inflow of gas to the TRML.

    \item[4.] In \autoref{sec:discussion} we argue that both previously proposed models to explain the $\edot/\vtL \propto \Da^{1/4}$ regime are inconsistent with the above observations. We then use the above observations to derive a new relation (\autoref{eq:scaling2}) which self-consistently recovers $\edot/\vtL \propto \Da^{1/4}$ assuming the excess fractal dimension of the interface is $d=1/2$ on small scales and with clear predictions of deviations from that scaling if $d \neq 1/2$.
\end{itemize}

\acknowledgments

The authors would like to thank Eliot Quataert, Eve C. Ostriker, Brent Tan, S. Peng Oh, Max Gr\"{o}nke, Camillo De Lellis, Chang-Goo Kim, Amiel Sternberg, Shyam H. Menon, Alexander Mayer, for useful discussions which improved this work. We thank the anonymous referee for a careful reading of the text and useful feedback which improved this work.

The authors gratefully acknowledge the support of the Kavli Institute for Theoretical Physics's 2024 program on ``Turbulence in Astrophysical Environments,'' where this work was initially conceived, and that of the Aspen Center for Physics's program ``Toward a Holistic Understanding of the Multi-scale, Multiphase Circumgalactic Medium'' where this work was continued. This research was therefore supported in part by grant NSF PHY-2309135 to the Kavli Institute for Theoretical Physics and NSF PHY-2210452 to the Aspen Center for Physics. L.L. acknowledges the support of the Simons Foundation under grant 965367.
D.B.F. gratefully acknowledges support from NSF through grants AST-2407387 and from NASA through grants HST-AR-17859.015-A and HST-AR-17559.009-A. This work was supported by a grant from the Simons Foundation (Grant Award ID BD-Targeted-00017375, DBF).
R.M. is supported by National Science Foundation (NSF) grants AST-2107872 and AST-2509269.
GLB acknowledges support from the NSF (AST-2108470, AST-2307419), NASA TCAN award 80NSSC21K1053, and the Simons Foundation through the Learning the Universe Collaboration.

The simulations presented in this work and much subsequent analysis was performed on the Flatiron Institute's \texttt{rusty} computing cluster.
The analysis presented in this article was performed in part on computational resources managed and supported by Princeton Research Computing, a consortium of groups including the Princeton Institute for Computational Science and Engineering (PICSciE) and the Office of Information Technology's High Performance Computing Center and Visualization Laboratory at Princeton University.
This research used both the DeltaAI advanced computing and data resource, which is supported by the NSF (award OAC 2320345) and the State of Illinois, and the Delta advanced computing and data resource which is supported by the NSF (award OAC 2005572) and the State of Illinois. Delta and DeltaAI are joint efforts of the University of Illinois Urbana-Champaign and its National Center for Supercomputing Applications.
\software{
{\tt scipy} \citep{scipy},
{\tt numpy} \citep{harrisNumpy2020}, 
{\tt matplotlib} \citep{matplotlib_hunter07},
{\tt adstex} (\url{https://github.com/yymao/adstex})
}

\newpage

\appendix

\begin{figure*}
    \centering
    \includegraphics[width=\textwidth]{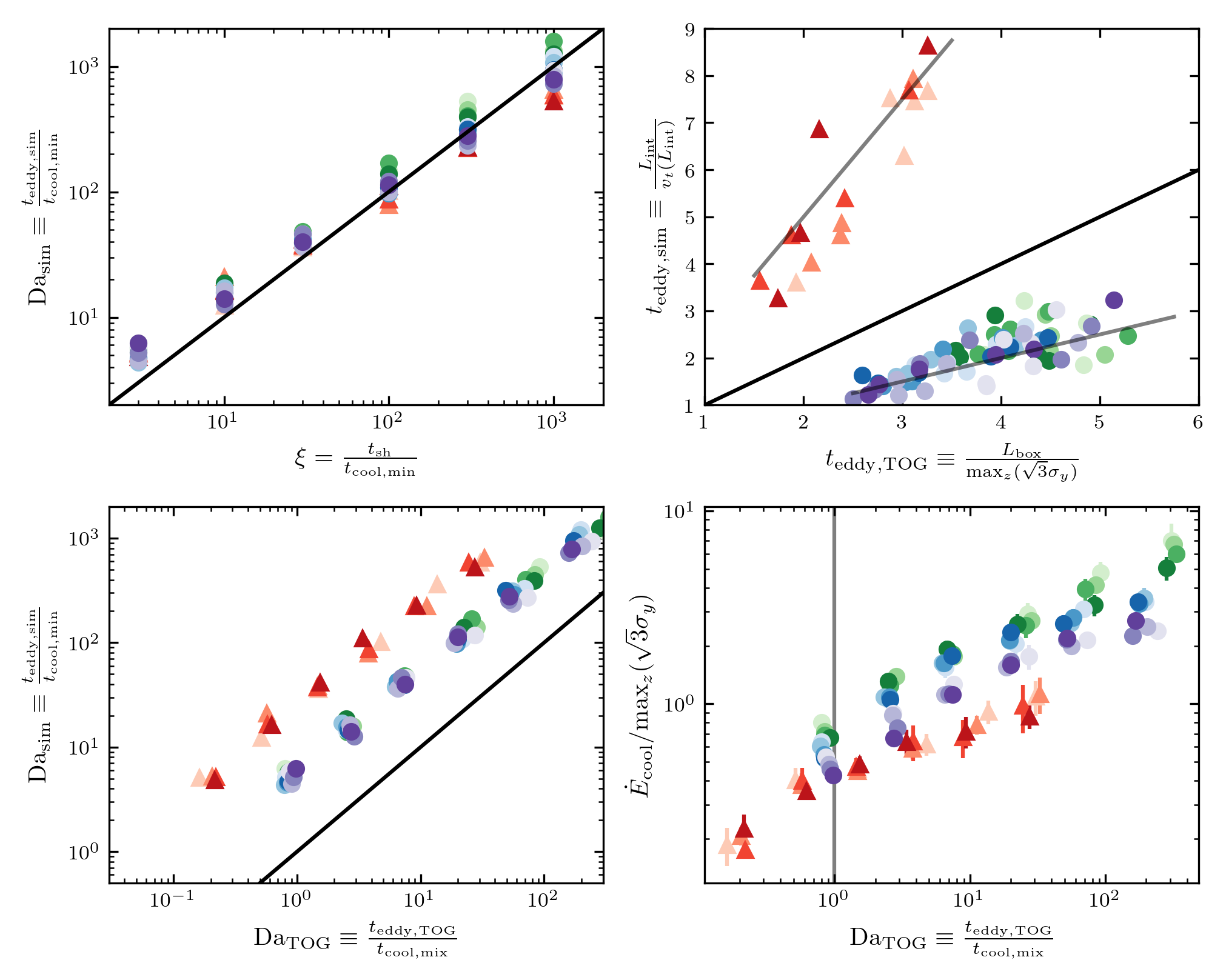}
    \caption{We compare various different quantities that are used for quantifying the importance of mixing relative to cooling. \textit{Top left}: $\Dasim$ as used in the main text (\autoref{eq:Da_measure}) relative to $\xi$ (\autoref{eq:xi_def}). \textit{Top right}: the eddy turnover time as measured in our simulations compared to how it is measured in \citet{Tan21}, $t_{\rm eddy, TOG}$. Gray lines indicate linear relationships, which are not consistent across simulations with varying $\mach$. \textit{Bottom left}: The measurement of $\Da$ used in this work compared to that used in \citet{Tan21}. \textit{Bottom right}: The equivalent of the top panel of \autoref{fig:vdiff_AT} for the measurement techniques of \citet{Tan21}. Black lines represent one-to-one relations in all panels in which they appear. In all panels triangles (circles) represent $\mach = 1/8\,\,\, (1/2)$ simulations, darker points represent higher resolution, and green, red, blue, and purple points represent $\chi = 30,\, 10^2,\, 10^2,\, 10^3$ simulations (red triangles are used here for the $\mach=1/8$, $\chi = 10^2$ simulations for consistency with \paperi).}
    \label{fig:Da_xi_comp}
\end{figure*}

\section{On the Definition of $\Da$ and its Relation to $\xi$}
\label{app:Da_def}

In \autoref{sec:intro} we define the Damk\"{o}hler number as
\begin{equation}
    \label{eq:Da_app}
    \Da \equiv \frac{\teddy(\Lint)}{\treact} = \frac{\Lint}{\vtL \treact} \, ,
\end{equation}
where $\Lint$ is the integral scale of the turbulence. For a given system, there is some freedom to choose exactly how to measure each of these quantities, and this is not done consistently across the literature. Different choices in how this measurement is done can result in large differences in $\Da$, to the extent that one set of choices would imply a simulation is in the fast cooling regime ($\Da > 1$) while another would imply that it is in the slow cooling regime ($\Da < 1$). In this appendix, we demonstrate this by comparing different choices from the literature and attempting to justify our choices, as well as relating $\Da$ to $\xi$ for the simulations presented here. These comparisons are summarized in \autoref{fig:Da_xi_comp}.

$\Da$ is a quantification of the importance of turbulence relative to cooling while $\xi$ compares the time for a parcel of gas to cross the box to the minimum cooling time in our simulations. We can control $\xi$ in our simulations by directly setting the shear velocity ($\vrel$) and cooling time ($\tcoolmin$), whereas $\Da$ arises naturally from the turbulence that develops in the system. Nevertheless, on comparing our measured $\Dasim$ to $\xi$ in the top left panel of \autoref{fig:Da_xi_comp} we see that they are nearly equal, especially in the $\mach = 1/2$ simulations (circles). At high $\xi$ in the $\mach = 1/8$ simulations, a larger fraction of the turbulent driving is provided by the inflow to the layer (bottom panel of \autoref{fig:vdiff_AT}) so that $\Dasim$ becomes somewhat smaller than $\xi$.

As stated in the main text, we measure $\Da$ in our simulations by computing the 2nd-order structure function of the $y$-velocity field and taking $\vtl = \sqrt{3}{\rm SF}_2(v_y)$ under the assumption that the $v_y$ is the only independent tracer of the turbulence across all simulated regimes (see Appendix B of \paperi). We then take the integral scale of the turbulence to be the scale that maximizes $\vtl$ take the eddy-turnover time of integral scale eddies as
\begin{equation}
    \teddy(\Lint) = \frac{\Lint}{\vtL} \, .
\end{equation}

In the work of \citet{Tan21} (as well as \citet{Das24,TMG25}) the turbulent velocity, $\vtL$ (which they refer to as $u'$), is calculated by computing the standard deviation of $v_y$ in constant $z$ slices of the domain: computing $\sigma_y(z)$. They then take $\vtL = \sqrt{3} \,{\rm max}_{z}\left( \sigma_y\right)$, the maximum value that this profile achieves along with the same $\sqrt{3}$ factor to account for tracing only one component of the turbulence. The method of \citet{Tan21} then takes the integral scale to be the lateral extent of the box so that the eddy-turnover time is
\begin{equation}
    t_{\rm eddy, TOG}(\Lint) = \frac{\Lbox}{\sqrt{3}\,{\rm max}_z(\sigma_y)} \, .
\end{equation}
This measurement has the advantage that it does not require computing structure functions, which can be computationally expensive. In the upper-right panel of \autoref{fig:Da_xi_comp} we compare these two measurements of the eddy-turnover time. We see that while they are linearly related to one another for the simulations probed here, this linear relation is not consistent across simulations with different mach numbers (triangles vs. circles).

The other key time-scale in estimating $\Da$ is the reaction or cooling time. We choose to use the minimum cooling time achieved in the layer assuming that it is iso-baric, $\tcoolmin$, which is a parameter of the simulations (see \autoref{sec:methods} and Section 3.2 of \paperi). \citet{Tan21} instead use the cooling time of `mixed' gas which they define as
\begin{equation}
    \label{eq:tcool_mix}
    \tmix = \frac{1}{\gamma -1} \frac{P_0}{\edotc(P_0, \Tmix)}
\end{equation}
where we use the cooling function defined in Section 3.2 of \paperi\ and $\Tmix \equiv \sqrt{\Tcold\Thot}$, the geometric mean of the hot and cold gas temperatures. We interrogate the meaning of this choice of $\tcool$ in \autoref{app:tcool}.

For our fiducial choice of cooling function shape (i.e. for all simulations in the main text) we have $\tmix/\tcoolmin \approx 10$. The choice of $\tmix$ or $\tcoolmin$ for $\treact$ in \autoref{eq:Da_app} can then clearly make a large difference in the inferred $\Dasim$. In the bottom left panel of \autoref{fig:Da_xi_comp} we compare our estimate of $\Dasim$ that we use to that used in \citet{Tan21}, which we call $\Da_{\rm TOG}$. This combines the estimate for $\teddy(\Lint)$ made above with $\tmix$. We see that in general $\Da_{\rm TOG} < \Dasim$ especially in the low $\mach$ simulations.

In the bottom right panel of \autoref{fig:Da_xi_comp} we show a version of the top panel of \autoref{fig:vdiff_AT} for the values of $\vtL$ and $\Da$ used in \citet{Tan21}. We see that for these choices the transition in scaling behavior from $\edot\propto\Da^{1/2}$ to $\edot\propto\Da^{1/4}$ happens much closer to $\Da \approx 1$, as was noted in \citet{FieldingFractal20} and \citet{Tan21}.

\begin{figure*}
    \centering
    \includegraphics[width=\textwidth]{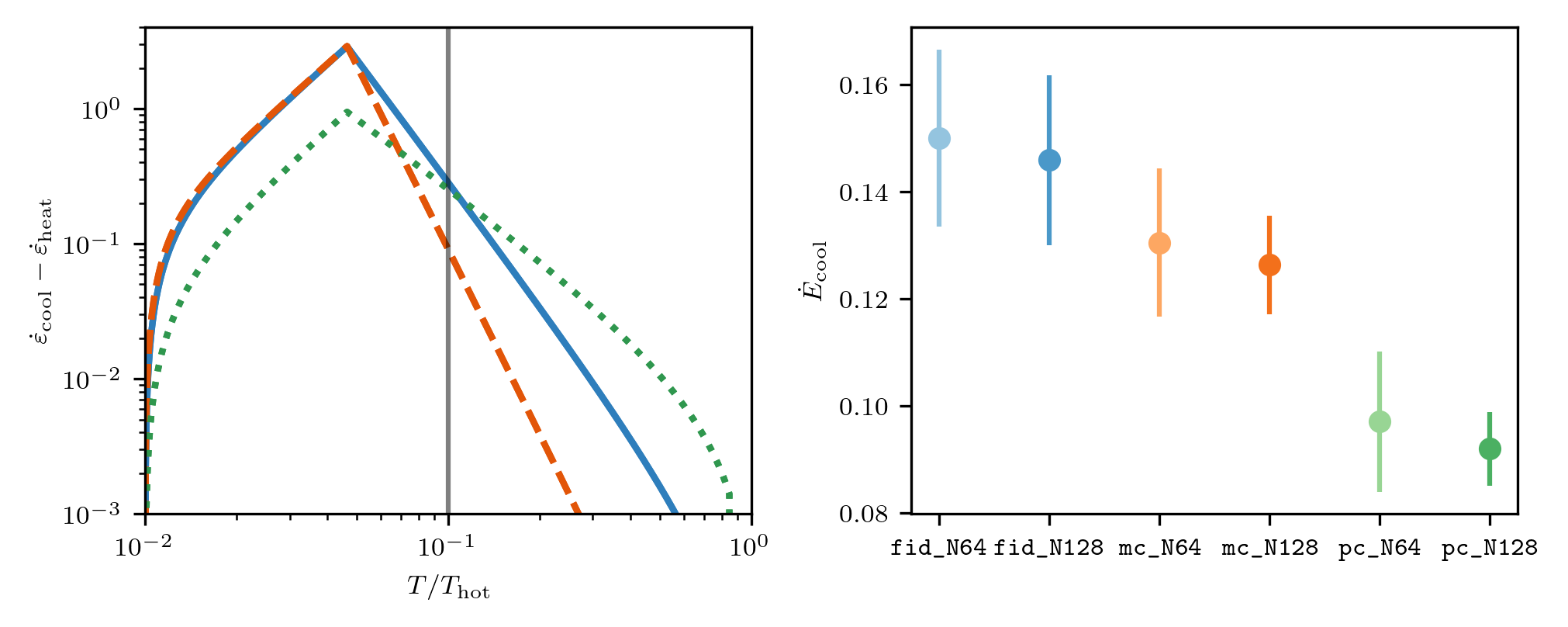}
    \caption{We compare the results of simulations with three different cooling functions. \textit{Left panel}: The different cooling functions investigated, the $\xi = 3$ fiducial cooling function for the simulation presented in the main text (blue), the same function but with $\bhi = 4.5$ (orange dashed) so that $\tmix$ is approximately 3 times the fiducial value, and the $\xi = 1$ cooling function but with $\bhi = 5/3$ (green dotted) so that $\tcoolmin$ is three times longer but $\tmix$ is the same as the fiducial simulations. \textit{Right panel}: The total cooling resulting from fiducial simulations (\texttt{fid}, blue), simulations which modify the mixed temperature cooling (\texttt{mc}, orange), and simulations which modify the peak temperature cooling (\texttt{pc}, green) at $\nres = 64\, \& \, 128$ (lighter and darker points respectively).}
    \label{fig:cf_comp}
\end{figure*}

\section{On the correct choice of $\tcool$}
\label{app:tcool}

The choice of $\Tmix$ for the temperature at which to measure the cooling time is based on the following argument from \citet{BegelmanFabian90}. In a TRML with turbulent velocity $\vtL$ at integral scale $\Lint$, the mass flux of hot gas into the mixing layer can be written as $\dot{m}_h \approx \eta_h \rho_{\rm hot} \vtL$ for dimensionless efficiency factor, $\eta_h$. If one assumes that the mass flux of cold gas into the mixing layer is determined by the Kelvin-Helmholtz instability on scale $l_c$ then the mass flux of cold gas to the layer can be written as $\dot{m}_c \approx \eta_c \rho_c l_c/t_{\rm KH}(l_c)$. Since\footnote{Truly, we should use $v_t(l_c)$ in this relation, but that is not how it is used in the original argument.} $t_{\rm KH}(l_c)= \sqrt{\rho_c/\rho_h}\, l_c/\vtL$ the cold gas mass flux in this argument becomes $\dot{m}_c \approx \eta_c \sqrt{\rho_c\rho_h} \vtL$, independent of $l_c$. One can then get an estimate of the temperature of the mixed gas by computing a mass-flux-weighted average:
\begin{equation}
    \label{eq:Tmix_BF90}
    \overline{T} = \frac{\dot{m}_c \Tcold + \dot{m}_h \Thot}{\dot{m}_{c} + \dot{m}_h}
    \approx \left[\frac{\eta_c + \eta_h\sqrt{\chi} }{\eta_h + \eta_c \sqrt{\chi}} \right] \Tmix \, ,
\end{equation}
so that $\overline{T} \approx \Tmix$ assuming the term in brackets is order unity.

There are several issues with the above argument: (i) there is no reason \textit{a priori} that $t_{\rm KH}$ should be used to estimate the flux of gas from the cold phase and not the hot phase, switching this assumption leads to $\overline{T} \approx \Tcold$ (in the limit $\chi \gg 1$) which is quite different, (ii) it is not clear why the time-scale for a linear instability ($t_{\rm KH}$) should be applied in the context of fully non-linear turbulent mixing, (iii) indeed, if one assumed a different time-scale associated with a different instability that may be more relevant in a given scenario (such as the Darrieus-Landau \citep{KK13}, Vishniac \citep{Vishniac83}, or Rayleigh-Taylor \citep{ChandrasekharBook} instabilities) one would arrive at a different formula for $\overline{T}$, and (iv) this argument only considers the role of turbulent diffusion (mixing) in estimating what the `typical' temperature of mixed gas is, while cooling is also generally very important in changing the temperature of a fluid element.

In order to test whether or not cooling at $\Tmix$ is truly indicative of cooling in the layer in the fully non-linear setting, we run a series of additional simulations with $\chi = 10^2$ and $\mach = 1/2$ but with modified cooling functions. These cooling functions are shown in the left panel of \autoref{fig:cf_comp}. In particular, we compare the $\xi = 3$ simulations with the fiducial cooling function (blue) with simulations which are identical except for $\bhi = 4.5$ (orange dashed), and $\xi = 1$ and $\bhi = 5/3$ (green dotted). We refer to the first modification as \texttt{mc} as it modifies the mixed temperature ($\Tmix$) cooling time by a factor of 3, and the second modification as \texttt{pc} as it changes the peak temperature ($\Tpk$) cooling time by a factor of 3 but keeps the cooling time at $\Tmix$ unchanged.

The total cooling rates for simulations at resolution of $\nres = 64\,\, \& \,\, 128$ are shown in the right panel of \autoref{fig:cf_comp}. As we can see, large changes in the cooling at $\Tmix$ result in relatively small changes in the total cooling rate (comparing orange to blue points, they are consistent within errors) whereas a change to the peak cooling temperature that keeps the cooling at $\Tmix$ unchanged results in significant differences (comparing blue to green points). We conclude that $\tcoolmin$ is a more meaningful time-scale on which to measure the rate of cooling in the layer than $\tmix$.

\bibliography{bibliography}{}
\bibliographystyle{aasjournal}

\end{document}